\documentclass{iaus}
\usepackage{graphicx}
\usepackage{natbib}
\usepackage{url}

\title[The Magellanic X-ray sources] 
{The Magellanic system X-ray sources}

\author[Gosling, A.~J. et al.]  
       {Andrew J. Gosling$^1$, Sean A. Farrell$^2$, Natalie A.
         Webb$^2$ \and Jari J.~E. Kajava$^1$}

\affiliation{$^1$Astronomy Division, Department of Physical Sciences,
  \\ P.O. Box 3000, University of Oulu, 90014, Finland
  \texttt{andrew.gosling@oulu.fi}\\ [\affilskip] $^2$Centre d’Etude
  Spatiale des Rayonnements, UPS/CNRS,\\ 9 Avenue du Colonel Roche,
  31028 Toulouse Cedex 04, France }

\pubyear{2008}
\volume{256}  
\pagerange{1}
\jname{The Magellanic System: Stars, Gas, and Galaxies}
\editors{Jacco Th. van Loon \& Joana M. Oliveira, eds.}
\begin{document}

\maketitle

\begin{abstract}

Using archival X-ray data from the second XMM-Newton serendipitous
source catalogue, we present comparative analysis of the overall
population of X-ray sources in the Large and Small Magellanic
Clouds. We see a difference between the characteristics of the
brighter sources in the two populations in the X-ray band. Utilising
flux measurements in different energy bands we are able to sort the
X-ray sources based on similarities to other previously identified and
classified objects. In this manner we are able to identify the
probable nature of some of the unknown objects, identifying a number
of possible X-ray binaries and Super Soft Sources.

\end{abstract}


\section{Introduction}

The Magellanic Clouds (MCs) provide an ideal location for the study of
source populations of all types.  At a distance of $48.5\,{\rm kpc}$
for the Large Magellanic Cloud (LMC; Macri et al., 2006) and $61\,{\rm
  kpc}$ for the Small Magellanic Cloud (SMC; Hilditch et al., 2005) it
is entirely possible to resolve individual components of stellar
populations with modern detectors.  For the study of X-ray sources
(XRS), there are a number of other features of these systems that are
beneficial.  Their comparatively small size on the sky means that
observations of the entire systems are feasible with a small number of
observations.  The low absorption towards the MCs, ${\rm n_H} = 1.5
\times 10^{21}$\footnote{\url{
  http://heasarc.gsfc.nasa.gov/cgi-bin/Tools/w3nh/w3nh.pl}} towards the
LMC and ${\rm n_H} = 3.0 \times
10^{21}$\footnotemark[\value{footnote}] towards the SMC benefits
observations of sources, especially those which emit the bulk of their
flux in the $0.1-3\,{\rm keV}$ soft band.  The Large Magellanic Cloud has a
metallicity of about forty per cent of the Milky Way, $Z =
0.0091\pm0.0007$, while the Small Magellanic Cloud has a metallicity
of about ten per cent of the Milky Way, $Z = 0.0050\pm0.0005$ (Keller
\& Wood, 2006) and their numerous epochs of star formation mean they
are ideal laboratories for exploring the formation processes and
evolution of stellar populations.

Prior to the launch of XMM-Newton and Chandra, the most complete X-ray
surveys of the MCs were performed by the ROSAT observatory (Haberl \&
Pietsch 1999; Haberl et al. 2000). These surveys detected $\sim\!750$
and $\sim\!500$ sources in the LMC and SMC respectively.  The improved
sensitivity and spatial resolution of the XMM-Newton observatory
(Jansen et al. 2001) means that with a comparatively small number of
observations it has already increased the number of XRS in the region
of the MCs threefold (see Section \ref{s:sources}) without approaching
full coverage of the MC fields.  In addition, there is a vast array of
datasets at other wavelengths for the MCs making follow-up of
identified XRS easier without the need to await targeted observations
to identify and analyse counterparts.

\section{The sources}
\label{s:sources}

We selected the sources from the second XMM-Newton serendipitous
source catalogue (2XMM; Watson et al. 2008) that were coincident with
the positions and extent of the LMC and SMC. The 2XMM catalogue is the
largest X-ray source catalogue ever produced, containing almost twice
as many discrete sources as previous all-sky surveys or other pointed
catalogues, with preliminary studies indicating that approximately
35\% of the real sources in the catalogue have not previously been
identified (Farrell et al. \textit{in prep.}). The catalogue provides
an effective dataset for generating large, well-defined samples of
various types of astrophysical object. The large sky area covered by
the serendipitous survey also means that 2XMM is a rich resource for
exploring the variety of the X-ray source populations and identifying
new examples of rare sources.

We used a strict selection criteria from the catalogue to discard
spurious detections and multiple detections of a single source. We
only selected those sources which had a {\tt SUM\_FLAG}$ < 3$ (an
automated pipeline flag indicating likely real sources), or with the
{\tt EPIC\_FLAG12} value set to True (a flag for sources which were
flagged as potentially spurious by the pipeline software but which
were determined to be real after manual inspection). In 2XMM there are
45 fields (a total of 208 observations) in the region of the LMC
containing 2069 discrete X-ray sources, and 12 fields (a total of 59
observations) in the region of the SMC containing 1204 discrete X-ray
sources.

\begin{figure}
  \begin{center}
    \includegraphics[width=0.51\textwidth]{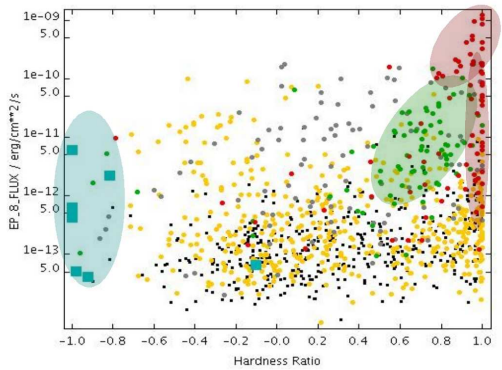}
    \caption[]{Hardness ratio vs. EPIC total band 8 ($0.2-12\,{\rm
        keV}$) flux, X-ray colour-magnitude diagrams (XCMD, see text
      for details) of identified variable XRSs in 2XMM. XRBs are
      marked in red, CVs in green, AGN in grey, stars in yellow and
      SSSs as turquoise squares. The figure shows selected regions of
      XRB, CV and SSS populations (see Section \ref{s:id} for
      details).}
    \label{f:pops}
  \end{center}
\end{figure}

\section{Identifying X-ray source types}
\label{s:id}

Figure \ref{f:pops} shows comparative flux diagrams used to identify
differences in XRS populations for a selection of identified XRSs from
2XMM variable source studies (Farrell et al. \textit{in prep.}). The
two panels compare the Hardness Ratio (HR) to the total (band 8) EPIC
flux (European Photon Imaging Camera; Jansen et al. 2001), hereby
referred to as an X-ray colour-magnitude diagram (XCMD). HR is defined
as $hard-soft/soft+hard$ , where $soft$ is the EPIC low energy flux:
$0.2-1\,{\rm keV}$, and $hard$ is the EPIC high energy flux:
$1-12\,{\rm keV}$.  Total EPIC flux is measured in the $0.2 - 12\,{\rm
  keV}$ energy band.  As can be seen in Figure \ref{f:pops}, sources
of similar type are found in specific regions of the XCMD.

Based on the positions of the different sources in Figure
\ref{f:pops}, we have selected X-ray colour and magnitude properties
that correspond to the different source populations. The two red
regions correspond to the bright, hard sources, mostly X-ray binaries
(XRBs). Close to this region and with some overlap is the green region
which contains fainter, softer sources such as Cataclysmic Variables
(CVs). Super Soft Source (SSSs; turquoise region), predominantly
thought to be white dwarfs undergoing steady thermonuclear burning
(Kahabka \& van den Heuvel 2006), are relatively faint in the
$0.2-12\,{\rm keV}$ range but have extremely soft colours.  Each
indicated region contains at least 50\% of the proposed source type
for that region.

\begin{figure}
  \begin{center}
    \includegraphics[width=0.49\textwidth]{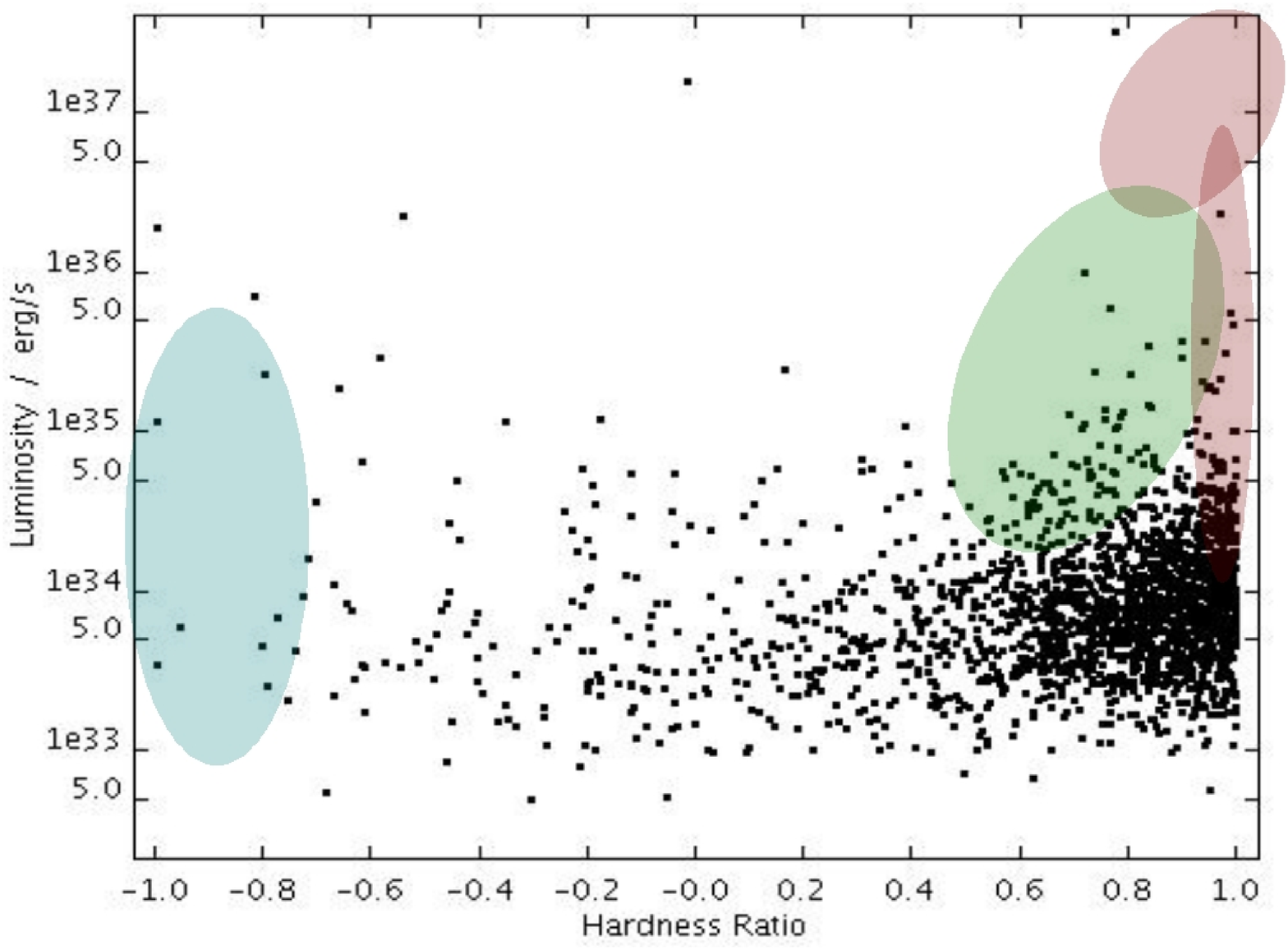}
    \includegraphics[width=0.49\textwidth]{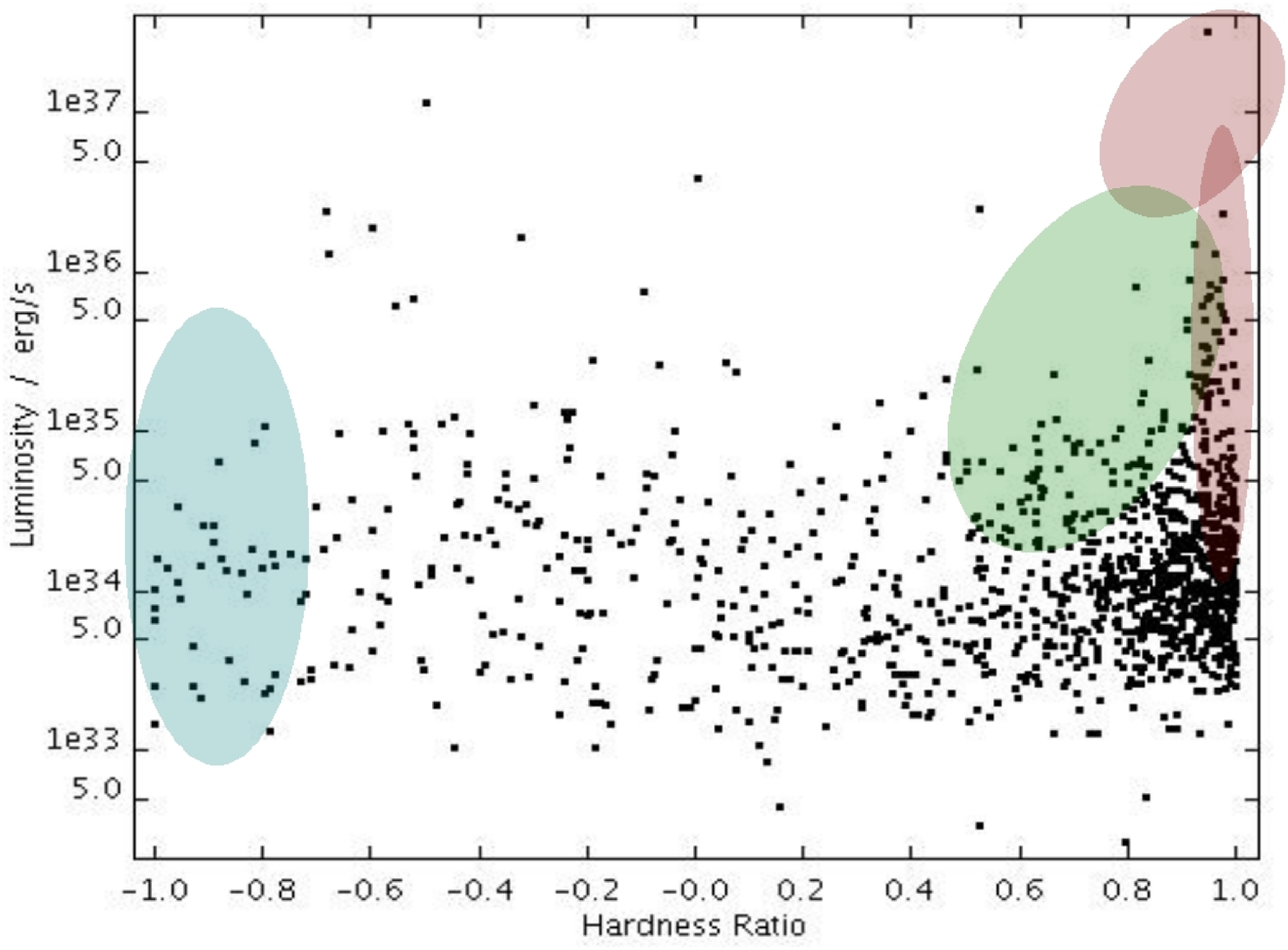}
    \caption[]{ XCMDs (Hardness ratio vs. Luminosity in the
      $0.2-12\,{\rm keV}$ band) for the LMC (left) and the SMC
      (right). These show the different XRS population locations from
      Figure \ref{f:pops} overlaid and scaled to the Magellanic Cloud
      XRS populations.  From these we have identified up to 20 new,
      unidentified sources of interest (with no listings in SIMBAD) in
      the first stages of our work.}
    \label{f:selection}
  \end{center}
\end{figure}

\section{The Magellanic X-ray source population}

By using the localisation of the different identified source
populations shown in Figure \ref{f:pops}, we can suggest the X-ray
type of the unidentified X-ray sources in the MCs, see Figure
\ref{f:selection}.  The LMC appears to have less sources that exhibit
extremes of HR.  The proportion of SMC sources that are bright and
soft is higher than in the LMC despite the absorption towards the LMC
being approximately half that towards the SMC\footnote{\url{
  http://heasarc.gsfc.nasa.gov/cgi-bin/Tools/w3nh/w3nh.pl}}.  However,
as we have exploited a catalogue composed of sources from requested
pointings of preferred regions and proportionally more of the SMC has
been covered than the LMC, our results could suffer from selection
effects.  Further coverage of both MCs will be required to verify
this.  We also note that the SMC contains a higher fraction of sources
that fall into the region dominated by XRBs and CVs than the LMC.
Comparison of the coordinates of the bright hard sources in the SMC to
the SIMBAD database shows that a significant fraction of the sources
are identified HMXBs, most likely to have formed in the most recent
epoch of star formation (Dray, 2006).  The fraction of the total
populations of the LMC and SMC that fall into the regions indicating
them to be likely XRBs appears to be smaller than the fraction of the
variable source population in these same regions in Figure
\ref{f:pops}.  This is partly due to selection effects; of the 7
brightest XRBs in the MCs, only 5 have been observed and of these only
1 is included in our population.  The others are flagged as potentiall
spurious by the pipeline software (i.e. {\tt SUM\_FLAG}$ > 2$) due to
numerous false detections surrounding them, a result of the
comparitively high flux of these XRBs.  It may also be linked to the
lower metallicities of the MCs. Low metallicity stars are known to
evolve quicker so the population of high mass stars in the MCs will be
short lived compared to similar mass stars in our Galaxy, reducing the
number of possible bright, hard, HMXB systems (Dray, 2006).  This may
have a secondary effect of increasing the number of LMXB systems in
the MCs as there will be a higher proportion of the compact remnants
of these high mass stars in the MCs.  This is one of the questions
that we hope to address in studying the total XRS population of the
MCs.

We have thus tried to determine the nature of some of the unknown
sources. We have identified two potential new SSSs in the LMC and one
potential new SSS in the SMC. We have identified eight possible new
XRBs in the LMC and four possible new XRBs in the SMC.  None of these
sources are coincident with identified sources in the SIMBAD database
or the associated literature.

\section{Conclusions and further work}

From the properties of identified sources in the 2XMM catalogue, we
have defined criteria to identify the likely nature of unidentified
X-ray sources in the LMC and SMC.  Using these properties, we have
outlined the differences in the two populations.  We show that in the
observed populations, there are proportionally more soft and bright
hard sources in the SMC than in the LMC.  Some of these features can
be attributed to selection effects.  However, it appears that others
may be the result of the different metallicities and star formation
histories of the two systems.

We have identified up to 20 new sources of interest in the MCs.  We
intend to perform full analysis and follow-up of all of these sources
including cross-matching of positions with optical and infrared
sources to identify candidate companions.

This work is in its preliminary stages and we expect to be able to
identify many new sources in the existing observations.  We have
already identified a number of sources as candidate X-ray Binaries,
Cataclysmic Variables and Super Soft Sources.  We also intend to carry
out further observations of the MCs to obtain a more complete
catalogue of all the X-ray sources.  \\

\footnotesize{
\noindent{\textit{Acknowledgements}}\\

AJG would like to thank Suomen Akatemia (The Finnish Academy) for
funding his post-doctoral position at Oulun Yliopisto (University of
Oulu) under grant number 110792.

This work is based on observations obtained with XMM-Newton, an ESA
science mission with instruments and contributions directly funded by
ESA Member States and NASA. The data was taken from the XMM-Newton
second serendipitous source catalogue. A large part of the source
selection was carried out using the TOPCAT package:
\texttt{http://www.starlink.ac.uk/topcat/}}


\end{document}